\documentclass[journal,12pt,onecolumn,draftclsnofoot,]{IEEEtran}
\usepackage{tikz}
\usepackage{fullpage}
\usepackage[colorlinks=true,linkcolor=black,citecolor=purple]{hyperref}
\usetikzlibrary{positioning,shadows}
\usepackage{breqn}
\usepackage{setspace}

\usepackage{algorithm}  

\usepackage{mathtools, cuted}
\usepackage{lipsum, color}
\usepackage{stfloats}
\usepackage{mathrsfs} 
\usepackage{amsmath} 
\usepackage{bm}
\usepackage{multicol}
\usepackage{amsmath,amssymb}
\usepackage{caption}
\usepackage{float}
\usepackage{setspace}
\usepackage{graphicx}
\usepackage{caption}
\usepackage[labelformat=simple]{subcaption}

\usepackage{tikz}
\usepackage{color}
\usepackage{amsmath} 
\usepackage{amssymb}
\usepackage{cite}
\usepackage{multicol}
 \usepackage{multirow}
 \usepackage{mathtools}

\usetikzlibrary{shapes,arrows,shadows}

\usepackage[keeplastbox]{flushend} 
\usepackage{array}
\newcolumntype{P}[1]{>{\centering\arraybackslash}p{#1}}

\begin{document}

\title{Perturbation Theory-Aided Learned Digital Back-Propagation Scheme for Optical Fiber Nonlinearity Compensation}
\author{Xiang Lin, Shenghang Luo, Sunish Kumar Orappanpara Soman, Octavia A. Dobre, Lutz Lampe, Deyuan Chang, Chuandong Li
\thanks{Xiang Lin and Octavia A. Dobre are with Faculty of Engineering and Applied Science, Memorial University, St. John's, NL, A1B 3X5, Canada; Shenghang Luo and Lutz Lampe are with Department of Electrical and Computer Engineering, University of British Columbia, Vancouver, BC V6T 1Z4, Canada; O. S. Sunish Kumar was with the Department of Electrical and Computer Engineering, the University of British Columbia, Vancouver, BC V6T 1Z4, Canada, and he is now with the School of Engineering, Ulster University, Jordanstown, BT37 0QB, UK; Deyuan Chang and Chuandong Li are with the Huawei Technologies Canada,
Ottawa, ON K2K 3J1, Canada. Corresponding email: xiang.lin@mun.ca}}

\maketitle
\vspace{-2.4cm}
\begin{abstract}
Derived from the regular perturbation treatment of the nonlinear Schrödinger equation, a machine learning-based scheme to mitigate the intra-channel optical fiber nonlinearity is proposed. Referred to as the perturbation theory-aided (PA) learned digital back propagation (LDBP), the proposed scheme constructs a deep neural network (DNN) in a way similar to the split-step Fourier method: linear and nonlinear operations alternate. Inspired by the perturbation analysis, the intra-channel cross phase modulation   term is conveniently represented by matrix operations in the DNN. The introduction of this term in each nonlinear operation considerably improves the performance, as well as enables the flexibility of PA-LDBP by adjusting the
 numbers of spans per step. The proposed scheme is evaluated by numerical simulations of a single-carrier optical fiber communication system operating at 32 Gbaud with 64-quadrature amplitude modulation  and 20 $\times$ 80 km transmission distance. The results show that the proposed scheme achieves approximately 3.5 dB, 1.8 dB, 1.4 dB, and 0.5 dB performance gain in terms of $\textit{Q}^2$ factor over the linear compensation, when the numbers of spans per step are 1, 2, 4, and 10, respectively. Two methods are proposed to reduce the complexity of PA-LDBP, i.e., pruning the number of perturbation coefficients and chromatic dispersion compensation in the frequency domain for multi-span per step cases. Investigation of the performance and complexity suggests that PA-LDBP attains  improved performance gains with reduced complexity when compared to LDBP in the cases of 4 and 10 spans per step.

\end{abstract}

\IEEEpeerreviewmaketitle


\section{Introduction}
Our digitally-interconnected world relies heavily on optical fiber communications  to transport the  majority of tremendous information. Over the past decades,  digital signal processing (DSP) has played an essential role to achieve reliable and high-speed  optical fiber communications \cite{savory2008digital,chang2016robust,zhang2017training,lin2018joint,winzer2018fiber,lin2019non}. To date, most of the linear impairments, such as chromatic dispersion (CD) and polarization mode dispersion, have been studied extensively and addressed well by DSP algorithms \cite{kuschnerov2009dsp,kikuchi2015fundamentals}. Fiber nonlinearity, on the other hand, appears to be the dominant barrier that limits the performance of today's optical fiber communication systems.  The power-dependent nature of the fiber nonlinear effect restricts the maximum launch power into the optical fiber, and hence, limits the effective signal-to-noise ratio (SNR) at the receiver. To mitigate the nonlinear effects, several DSP techniques have been proposed, such as the digital back-propagation (DBP) and perturbation theory-based (PB) nonlinearity compensation (NLC) \cite{ip2008coherent,liang2014multi, amari2017survey, liang2015correlated,liang2017perturbation}. DBP attempts to compensate the deterministic fiber nonlinearity at the receiver by emulating the signal propagation in the fiber channel in a reverse direction. Its complexity is significantly higher than linear compensation. PB-NLC is typically developed based on the first-order perturbation of the nonlinear Schrödinger equation (NLSE).  The PB-NLC technique exhibits reduced computational complexity in comparison with DBP.  A number of variants based on DBP and PB-NLC have been developed to strike a balance between performance and complexity \cite{secondini2016single,kumar2019enhanced,orappanparasoman2021second}. Nonetheless, the sophisticated nature of the distortion caused by the interaction between the CD, fiber nonlinearity, and amplified spontaneous emission noise have made it challenging to develop a NLC DSP algorithm to achieve good performance and low complexity suitable for commercial implementation.

Recently, the potential of machine learning to overcome fiber nonliearity  has been explored in the literature \cite{zhang2018non,amari2019fiber,amari2019machine,sidelnikov2018equalization,zhang2019field}.
In general, these techniques seek to establish a mapping between the input  and  output through training, and the learned mapping is employed at the inference stage to remove nonlinear effects. For example, in learned PB-NLC, the data triplets are used as input, and the output is the additive nonlinear distortion field\cite{zhang2019field}.  
The mapping is obtained through training, using a large amount of data instead of rigorous mathematical reasoning; as such, it typically requires empirical experience to find out the model hyper-parameters \cite{goodfellow2016deep}. 
Unlike such ``black-box" solutions, a technique that employs a  deep neural network (DNN) to unfold the conventional DBP, referred to as the learned DBP (LDBP),  has been proposed recently \cite{hager2018nonlinear,hager2020physics}. LDBP interprets the split-step Fourier method (SSFM) as a DNN with each hidden layer as the linear section and the corresponding activation function as the nonlinear section.  As a result, LDBP incorporates the optimization capability of a DNN into the conventional DBP by parametrizing the SSFM and training the parameters through supervised learning.  Performance improvement and complexity reduction compared to the conventional DBP are achieved. Additionally, this technique leads to explicit hyper-parameters selection, such as the number of layers and the activation function.  
This approach has been also validated through experiments \cite{fan2020advancing,oliari2020revisiting}.  More recently, a convolutional neural network (CNN) has been considered to replace the DNN structure in LDBP \cite{sidelnikov2021advanced}. To summarize, the LDBP technique accomplishes the linear steps through the weight matrices operation in DNN or CNN, and the nonlinear steps through the nonlinear activation functions. 
 However, the number of layers in the DNN increases with the number of fiber spans in the link, and such a deep structure  faces several challenges, such as the extended training time/memory and the well-known vanishing gradient problem \cite{goodfellow2016deep}. 
 
In this paper, we propose a novel design, where the nonlinear steps of LDBP are improved with the aid of the first-order perturbation analysis. One may see that the proposed perturbation-aided (PA)-LDBP scheme shares some merits with the enhanced SSFM (ESSFM), where the nonlinearity compensation requires the adjacent symbols of the current symbol of interest. However, the differences between the proposed PA-LDBP and  ESSFM  in \cite{hager2020physics} and \cite{sidelnikov2021advanced} are evident.  Specifically, in \cite{hager2020physics}, the focus is on the performance gain of LDBP that adopts ESSFM over the conventional  DBP with ESSFM; while in \cite{sidelnikov2021advanced}, the aim is to improve the performance at each nonlinear step, and only 1 span per step is considered. In this work, we show that the benefits of our proposed PA-LDBP are fourfold:
\begin{itemize}
\item The perturbation analysis is employed to facilitate the DNN initialization. This physics-informed initialization guides the training deterministically towards a promising solution, and thus, significantly reduces the training effort. 
\item It provides a flexible structure which enables a multi-span per step configuration.
\item It outperforms LDBP for the same number of spans per step.
\item It achieves a reduction in complexity when multi-span per step is considered.
\end{itemize}
The remainder of this paper is organized as follows. In Section II,  the system model, the LDBP, and the proposed PA-LDBP scheme are introduced. In Section III, a comprehensive performance analysis is carried out, the performance of PA-LDBP is compared to that of LDBP, and the initialization for PA-LDBP is discussed. In Section IV, the complexity analysis is provided, and two approaches to reduce the overall complexity are presented. Finally, conclusions are drawn in Section V. 

Notations: Throughout the paper, the upper case bold letters
represent matrices, while the lower case bold font denotes column vectors. 
$\mathbf{A}^{T}$ indicates the transpose of matrix $\mathbf{A}$, and $\mathbf{A}^{*}$ denotes the conjugate of matrix $\mathbf{A}$.  $\mathbb{C}$ represents the complex-valued domain, and $\mathbb{Z}$ represents the integer domain.

\vspace{-0.4cm}
\section{The Proposed PA-LDBP Scheme}
\vspace{-0.4cm}
In this section, the system model is introduced, followed by a brief description of the LDBP technique. Then, details of the proposed PA-LDBP are provided.
\vspace{-0.6cm}
\subsection{System Model}
\vspace{-0.4cm}
In a single mode optical fiber, the propagation of the optical field envelope $u$ at the retarded time frame $t$ and distance $z$ is governed by the scalar NLSE, which is given as
\begin{equation}
\frac{\partial}{\partial  z}u(z,t)+ \frac{\alpha}{2}u(z,t) + j\frac{\beta_2}{2} \frac{\partial^2}{\partial t}u(z,t)  = j \gamma |u(z,t)|^2 u(z,t), 
\label{Eq:NLSE}
\end{equation}
where $\alpha$ is the attenuation coefficient, $\beta_2$ is the group velocity dispersion coefficient, and $\gamma$ is the nonlinearity coefficient \cite{agrawal2011nonlinear,ip2008compensation}.
The signal launched into the optical fiber channel is given by 
\begin{equation}
u(0,t)=\sqrt{P}\sum_{n=-\infty}^{\infty} s_n g(0,t-nT),
\end{equation}
where $P$ is the launch power, $s_n$ is the $n$th symbol, $g(0,t-nT)$ is the pulse, and $T$ is the symbol duration. The symbol sequence is assumed independent and identically distributed  and with unit power.  After propagating through the optical channel, the signal is coherently detected.  Then, it passes through a low-pass filter and is sampled at $t=nT, n\in\mathbb{Z}$. Subsequently, a sampled sequence, $\mathbf{x}=[x_1,  x_2, ..., x_{N}]^{T}$, is obtained.  
 The objective of NLC techniques is to recover information from $\mathbf{x}$.

\vspace{-0.6cm}
\subsection{The LDBP Technique}
\vspace{-0.25cm}
DBP can be viewed as a concatenation of linear and nonlinear steps. In the absence of noise, the transmitted signal is estimated by reversing the NLSE:
\begin{equation}
\frac{\partial}{\partial  z}u(z,t) = (\hat{D}^{-1}+\hat{N}^{-1})u(z,t),
\end{equation}
where $\hat{D}=-j \frac{\beta_2}{2}\frac{\partial ^2}{\partial t^2}-\frac{\alpha}{2}$ and $\hat{N}=j\gamma |u(z,t)|^2$ are the linear and nonlinear operators, respectively, and$(\cdot)^{-1}$ is the reverse operation \cite{ip2008compensation}. Given the sampled signal, $\mathbf{x}$, the linear step of DBP is expressed as
\begin{equation}
\mathbf{x}^{\text{CD}}=\exp (\frac{\alpha}{2}\mu )\mathbf{F}^{-1} ( \mathbf{Fx}  \exp (-j \frac{\beta_2}{2} \boldsymbol{\omega}^2 \mu) ),
\label{eq:CDcomp}
\end{equation}
where  $\mathbf{F}$ is the $N \times N$ discrete Fourier transform (DFT) matrix,  $\mathbf{F}^{-1}$ is the inverse DFT matrix,  $\mu$ is the step size, and $\boldsymbol{\omega}$ is the DFT angular frequency, whose $i$th element is given by $\omega_i=2\pi f_i$ ($f_i=f_s (i-1)/N$ if $i < N/2$ and $f_i=f_s (i-1-N)/N$ if $i \geq N/2$). $f_s$ is the sampling rate.
After that, a nonlinear operation is performed sample-by-sample in time domain according to
\begin{equation}
x_{n}^{\text{NL}} =x_{n}^{\text{CD}} \text{exp} \left(-j\zeta \gamma L_{\text{eff}}(\mu)  |x_{n}^{\text{CD}}|^2\right),
\label{eq:dbpnonlinearstep}
\end{equation}
where $x_{n}^{\text{NL}}$ is the $n$th sample after the nonlinear step, $\zeta$ is a parameter which needs to be optimized empirically, $L_{\text{eff}}(\mu)=(1-\exp(-\alpha \mu)/\alpha)$ is the effective nonlinear length, and $x_{n}^{\text{CD}}$ is the $n$th sample in $\mathbf{x}^{\text{CD}}$.  
Typically, a smaller step size leads to an enhanced performance but results in higher complexity. 

LDBP employs the connection between the DBP and a DNN: in both cases, linear and nonlinear steps alternate. 
In LDBP, all linear steps are parametrized as the DNN weight matrices  $\mathbf{W^{(1)}},...\mathbf{W}^{(\ell)},...\mathbf{W}^{\text(L)}$, where $\mathbf{W^{(\ell)}} \in \mathbb{C}$ is an $N$-by-$N$ matrix, the superscript $(\ell)$ represents the $\ell$th step, and $L$ represents the number of steps. 
On the other hand, the nonlinear operation at the $\ell$th step is performed by applying the activation function 
\begin{equation}
\sigma ^{(l)}(x_{n}^{\text{CD}})=x_n^{\text{CD}} \text{exp}(-j \eta^{(\ell)} \gamma L_{\text{eff}}(\mu) |x_n^{\text{CD}}|^2)
\end{equation}
in the DNN, where $\eta^{(\ell)}$ corresponds to $\zeta$ in \eqref{eq:dbpnonlinearstep} and is configured as a non-trainable parameter in LDBP \cite{hager2020physics}.
By concatenating all the steps,  LDBP can be expressed concisely as 
\begin{equation}
\Pi_{\text{LDBP}}(\mathbf{x})={\sigma}(\mathbf{W}^{(L)}...\sigma(\mathbf{W}^{(1)}\mathbf{x})).
\end{equation} 
A parameter set $(\mathbf{W}^{(1)},...,\mathbf{W}^{(\ell)},...,\mathbf{W}^{(L)})$ is defined, which is optimized through supervised learning. It is worth noting that the rows of the linear matrix $\mathbf{W}^{\textit(\ell)}$ are circularly-shifted versions of \\$\mathbf{h}^{(\ell)}=\left(h_{-V}^{(\ell)}, \ldots, h_{-1}^{(\ell)}, h_{0}^{(\ell)}, h_{1}^{(\ell)}, \ldots, h_{V}^{(\ell)}, 0, \ldots, 0\right)$, where $h_{v}^{(\ell)} \in \mathbb{C}, v=-V, \ldots, V$ represent the finite impulse response (FIR) filter coefficients for CD compensation \cite{hager2020physics}. Note that the filter is symmetric, so the trainable parameters at the $\ell$th step are $\mathbf{h_v}^{(\ell)}= \left(h_{0}^{(\ell)}, h_{1}^{(\ell)}, \ldots, h_{V}^{(\ell)}\right)$. Thus, the parameter set for the entire model is downsized to $\boldsymbol{\theta}=\left\{\mathbf{h_v}^{(1)}, \ldots, \mathbf{h_v}^{(\ell)}, \ldots, \mathbf{h_v}^{(L)}\right\}$.  
Recently, a CNN structure has been employed to replace the DNN structure \cite{sidelnikov2021advanced}. The CNN-based LDBP considers a block of symbols as the input, while the output is the equalized symbol corresponding to the center of the input block. In spite of different structures, both DNN and CNN perform  circular convolutions-based time-domain equalization (TDE) to compensate for the CD at each step. 

LDBP offers important advantages over other machine learning-based NLC methods which are data-driven \cite{amari2019fiber,zhang2019field}. Firstly, the DNN structure is based on physical principles, which is the SSFM approximation of the signal propagation over optical fibers. It follows that hyper-parameters, such as the type of the activation function and the number of layers, are associated with physical parameters and their choice or optimization are interpretable similar to the case of conventional DBP.  Secondly, the initialization of the DNN weights for linear steps  can be based on analytical results used in the DBP, which greatly facilitates successful and fast training. These benefits will be inherited by the proposed PA-LDBP scheme.

\vspace{-0.55cm}
\subsection{The PA-LDBP Scheme}
\vspace{-0.15cm}
LDBP has been shown to substantially improve the performance and complexity when compared to the conventional DBP \cite{hager2020physics}.  An important factor in the complexity reduction of LDBP is the pruning of network parameters, i.e., the number of non-zero elements of the weight matrices $\mathbf{W}^{(l)}$, which effectively reduces the number of coefficients of the TDE filter  \cite{hager2018deep}.  However, it is noticed that most of the investigations of the LDBP scheme consider self phase modulation (SPM), and thus, a relatively small number of spans per step is required.  Alternative speaking, the number of layers in LDBP needs to increase  with the number of fiber spans for successful nonlinearity compensation. Hence, A very deep DNN is required, which is challenging \cite{goodfellow2016deep}. 

\begin{figure*}[t]
\centering
\includegraphics[width=1\textwidth]{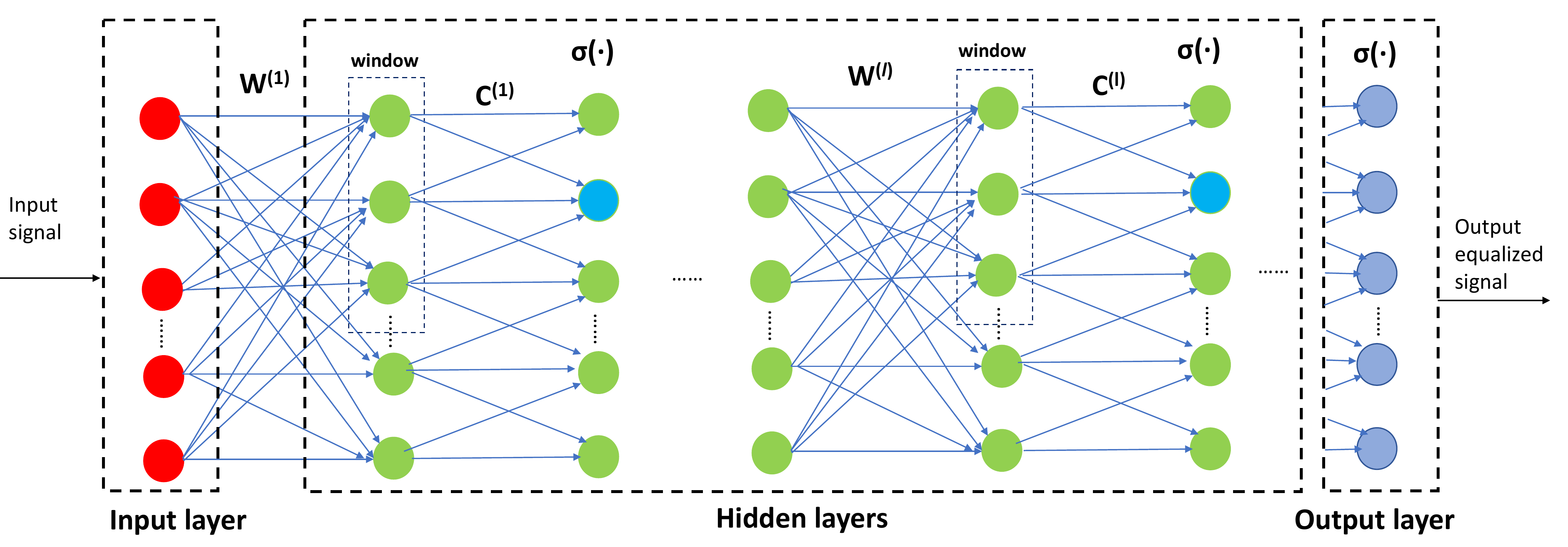}
\caption{PA-LDBP structure.}
\label{fig:LBPAstructure}
\vspace{-0.4cm}
\end{figure*}

To overcome the above-mentioned problems, we propose to refine the nonlinearity compensation by considering both SPM and intra-channel cross-phase modulation (IXPM). The details are presented in the following. 

Without loss of generality, we consider the first step of PA-LDBP, and  the first-order nonlinear distortion to the $n$th  sample  is given as 
\begin{equation}
\Delta_n=j P^{\frac{3}{2}} \sum_{m,k}  x^{\text{CD}}_{k} {(x^{\text{CD}}_{m+k})}^{\ast} x^{\text{CD}}_{m}  C_{m,k},
\label{Eq: perturbation distortion cal}
\end{equation}
where $C_{m,k}$ is the perturbation coefficient that can be obtained by \eqref{perturbationcoeff} shown at the bottom of the page, with $f(z)$ as the power profile function defined as $\text{exp}(-\alpha z)$ \cite{liang2014multi}, \cite{tao2014analytical}. It is worth noting that the distortion for other steps can be calculated in a similar way using \eqref{Eq: perturbation distortion cal}.
\begin{figure*}[b]
\begin{align}
C_{m,k}=\frac{1}{T} \int_{0}^{z} dz \gamma f(z) \int dt {g^{*}(z,t)}   g(z, t-mT) g(z,t-kT){g^{*}(z,t-(m+k)T)}
\label{perturbationcoeff}
\end{align}
\end{figure*}
By expanding \eqref{Eq: perturbation distortion cal} and performing some algebraic simplifications, the distortion field can be  separated into the SPM, IXPM, and intra-channel four-wave mixing (IFWM) effects, as follows:
\begin{equation}
\Delta_n=j P^{\frac{3}{2}} x_{n}^{\text{CD}} \left[ |x^{\text{CD}}_{n}|^2  C_{0,0} + 2\sum_{k \neq 0}|x^{\text{CD}}_{k}|^2 C_{0,k}  \right] \\
 + j P^{\frac{3}{2}} \sum_{m \neq 0, k \neq 0} x^{\text{CD}}_{k} {(x^{\text{CD}}_{m+k})}^{\ast} x^{\text{CD}}_{m}  C_{m,k}.
\label{Eq: perturbation distortion cal1}
\end{equation}

Let $\phi_{n}=P^{\frac{3}{2}}\left[ |x^{\text{CD}}_{n}|^2  C_{0,0} + 2\sum_{k \neq 0}|x^{\text{CD}}_{k}|^2 C_{0,k}\right]$; then, when SPM and IXPM are both considered, the nonlinear step in PA-LDBP can be represented as
\begin{align}
\sigma(x_{n}^{\text{CD}})=&x_{n}^{\text{CD}}-\Delta_n \nonumber\\
  \approx&x_{n}^{\text{CD}} (1-j\phi_{n}) \nonumber \\
  \approx & x_{n}^{\text{CD}} \exp(-j\phi_{n}) \nonumber \\
= &x_{n}^{\text{CD}} \exp \left( -j P^{\frac{3}{2}}   (\textbf{x}^{\text{CD}}_{k})^{T} \textbf{c}_{0}   \right), 
\label{eq:spmxpm}
\end{align}
where $\textbf{x}_{k}^{\text{CD}}$ is the vector  $[ |x_{n-k}^{\text{CD}}|^2,...,|x_{n}^{\text{CD}}|^2,...,|x_{n+k}^{\text{CD}}|^2]^T$ and $\mathbf{c_0}$ represents the perturbation coefficients vector which accounts for the SPM and IXPM effects, i.e., $[2C_{0,k},...,C_{0,0},...,2C_{0,k}]^T.$ Note that \eqref{eq:spmxpm} holds when $\phi_{n}\ll 1$. By  carefully examining \eqref{eq:spmxpm}, it can be seen that PA-LDBP involves one additional operation when compared to LDBP at each nonlinear step. This operation aims to learn the nonlinearity interaction between a number of adjacent samples. 
To generalize, we consider the $\ell$th step, which includes 2 sets of weights: $\mathbf{W}^{(\ell)}$ and $\mathbf{C}^{(\ell)}$. The former can be configured in the same way as in LDBP for CD compensation;  the rows of the latter are circularly-shifted versions of $\mathbf{c_0}^{(\ell)}$.  By concatenating all the steps, one can summarize the PA-LDBP as
\begin{equation}
\Pi_{\text{PA-LDBP}}(\mathbf{x})={\sigma}(\mathbf{C}^{(L)}\mathbf{W}^{(L)}...\sigma(\mathbf{C}^{(1)}\mathbf{W}^{(1)}\mathbf{x})).
\end{equation} 
The resulting structure of PA-LDBP is shown in Figure \ref{fig:LBPAstructure}. The dashed-line window, given as an example, includes three neurons for nonlinearity calculation contributing to the neuron (in blue) of the next layer. 
Although the complexity at each step increases compared to LDBP, the introduced extra operation significantly improves the nonlinearity compensation performance, and enables a flexible structure with multi-span per step.  Alternatively speaking, compared to LDBP,  PA-LDBP could achieve similar performance with a smaller number of steps, $L$. 

\begin{figure*}[t]
\centering
\includegraphics[width=1\textwidth]{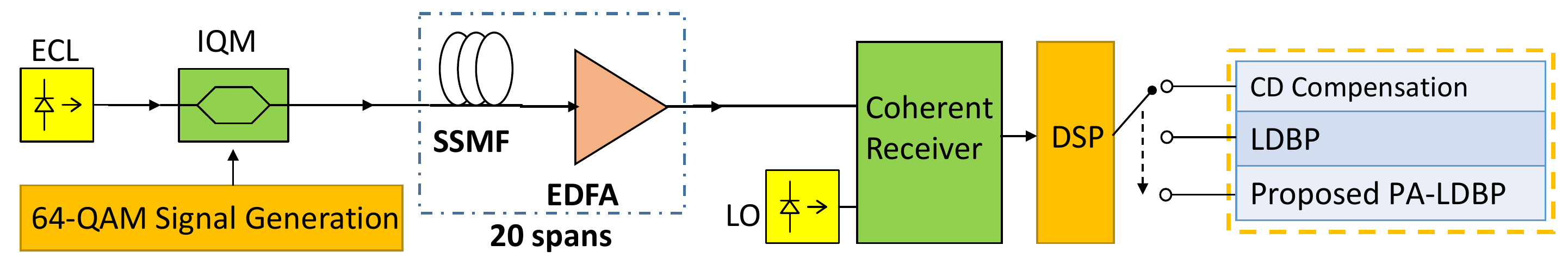}
\caption{Simulation system block diagram. LO: local oscillator.}
\label{fig:simulationsystem}
\vspace{-0.3cm}
\end{figure*}

\vspace{-0.3cm}
\section{Numerical Results and Discussions}
\vspace{-0.3cm}
In this section, the performance of the proposed PA-LDBP is investigated through numerical simulations. The details about the simulation implementation are provided, followed by the demonstration of the simulation results. 

\vspace{-0.5cm}
\subsection{Simulation Setup}
\vspace{-0.3cm}
\subsubsection{Optical Fiber System Setup}
A single carrier system, shown in Fig. \ref{fig:simulationsystem}, is employed to generate the training and testing data.  A 64-quadrature amplitude modulation (QAM) signal at 32 Gbaud with a roll-off factor of 0.1 is generated to modulate an optical carrier through the in-phase quadrature modulator (IQM). The carrier is provided by an external cavity laser (ECL) with 1550.12 nm center wavelength. Then, the modulated signal is fed into 20 spans of standard single mode fiber (SSMF). For each span, the SSMF is 80 km, followed by an Erbium-doped fiber amplifier (EDFA).  The SSFM has an attenuation coefficient of 0.2 dB/km, a dispersion parameter of 17 ps/nm/km, and a nonlinear coefficient  of 1.3 /W/km. The EDFA has a 5 dB noise figure and 16 dB gain. The propagation is emulated by the SSFM with 100 steps per span. 
Similar to \cite{hager2020physics} and \cite{sidelnikov2021advanced}, carrier frequency offset and laser phase noise are not considered in the simulation. 
At the receiver, the processing includes static equalization and phase rotation recovery. The $Q^2$ factor is used as performance metric, and is defined as $Q^2=20 \text{log}_{10}(\sqrt(10) \text{erfc}^{-1}(8\text{Ber}/9))$, where $\text{erfc}(\cdot)$ is the complementary error function and $\text{Ber}$ is the system bit error rate.

\begin{figure*}[t!]
\centering
\begin{subfigure}{0.33\textwidth}
  \centering
  \includegraphics[width=0.98\linewidth]{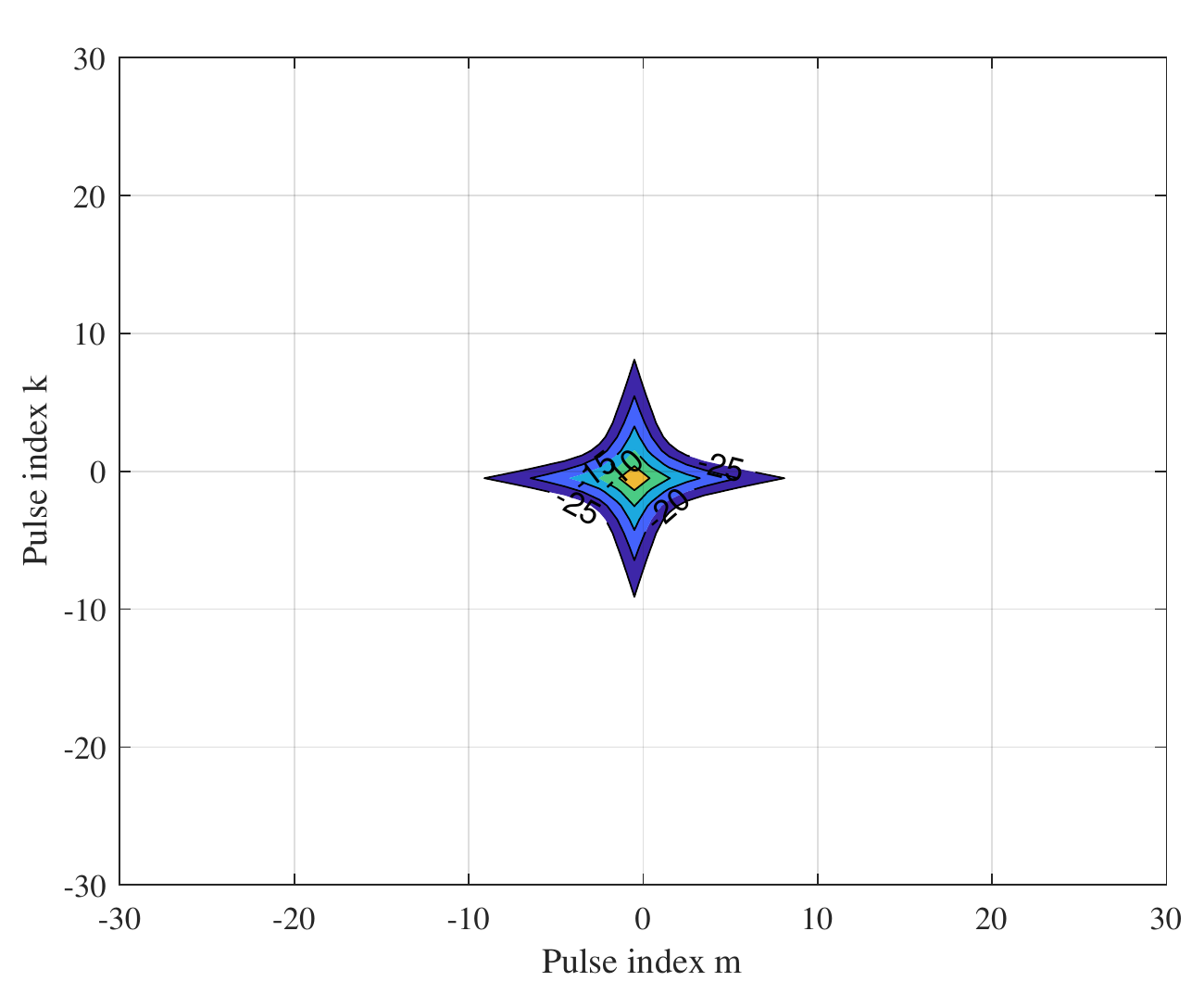}
  \caption{$C_{m,k}$ for 1 span}
\end{subfigure}%
\begin{subfigure}{0.33\textwidth}
  \centering
  \includegraphics[width=0.98\linewidth]{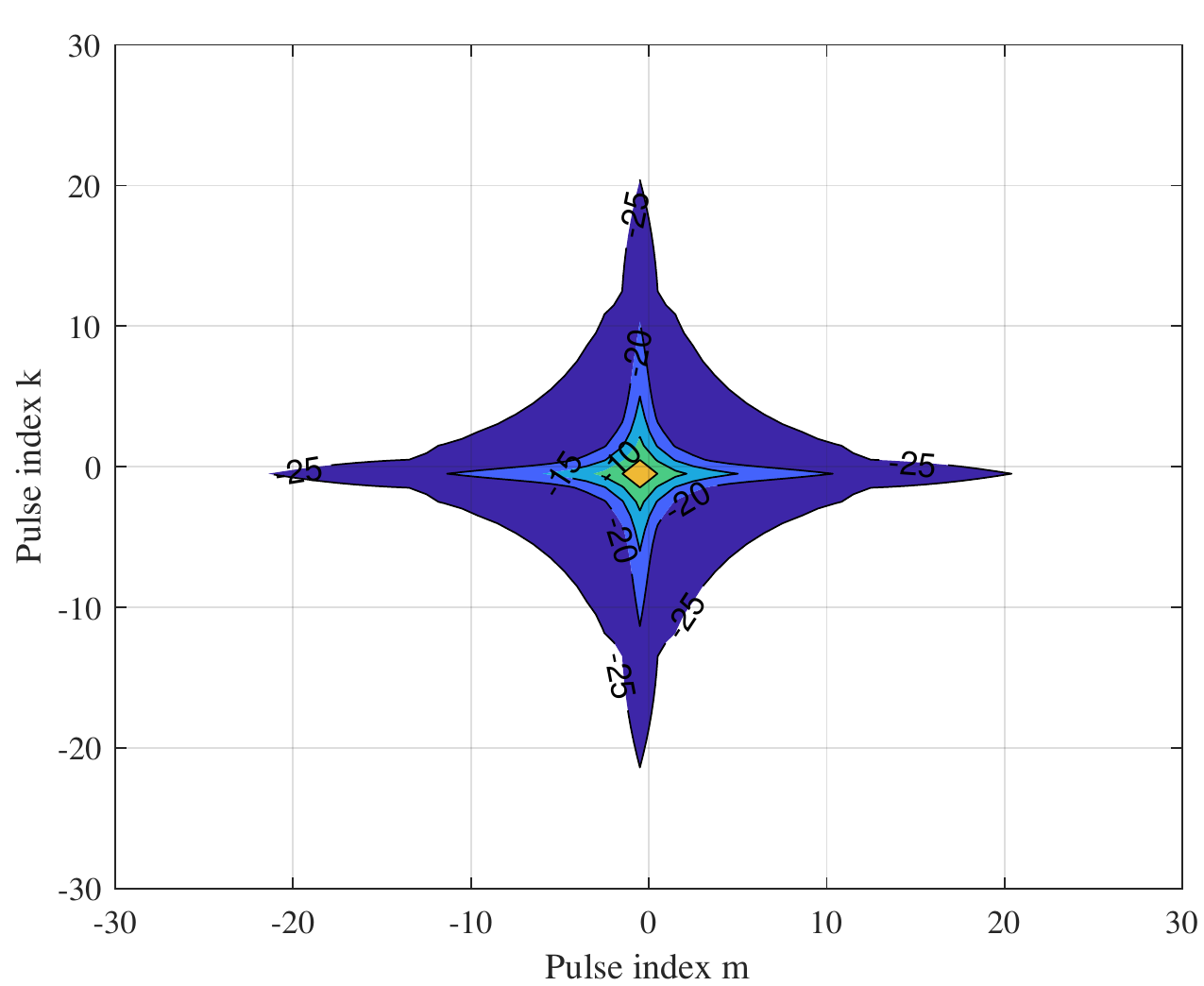}
  \caption{$C_{m,k}$ for 2 spans}
\end{subfigure}%
\begin{subfigure}{0.33\textwidth}
  \centering
  \includegraphics[width=0.98\linewidth]{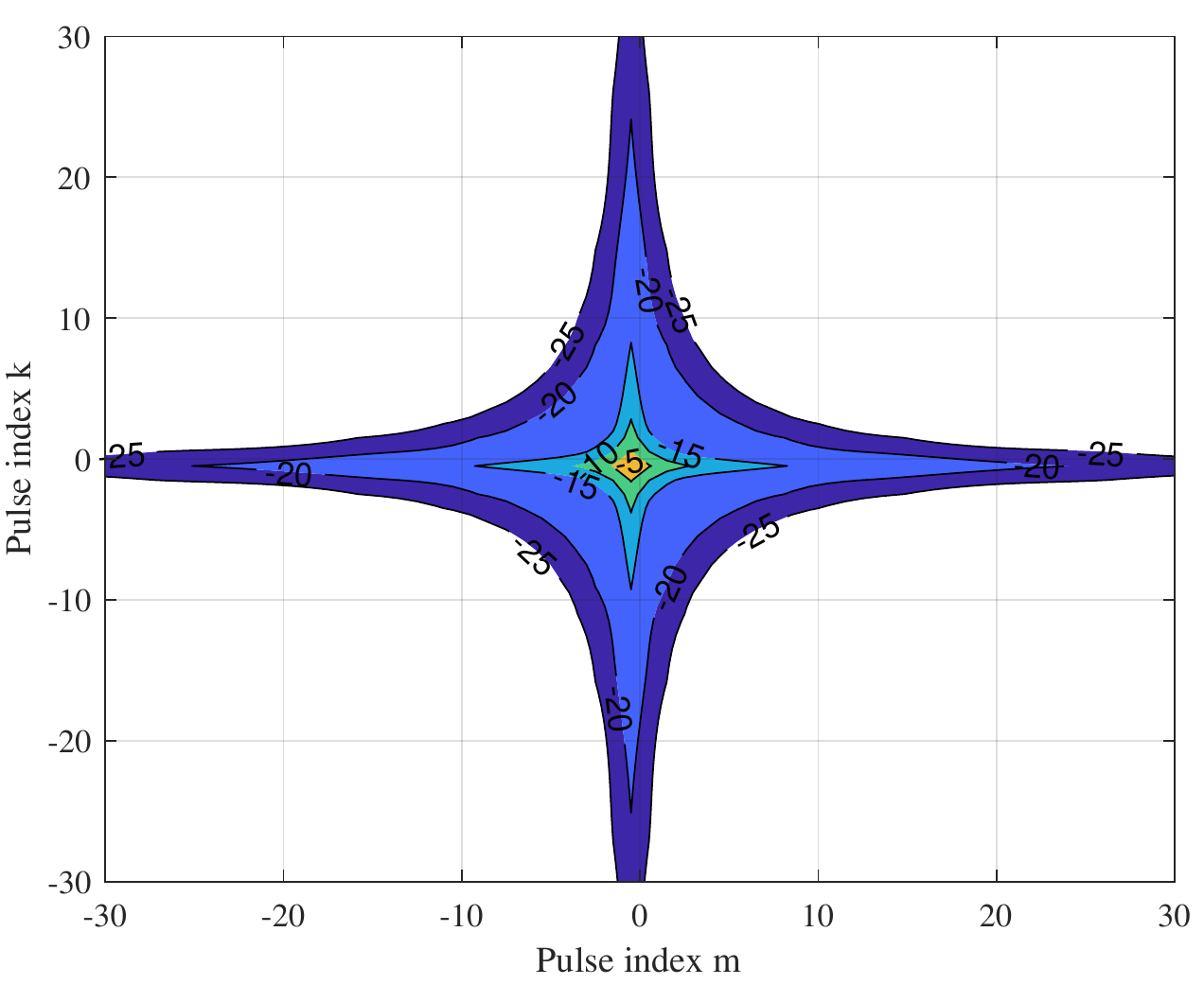}
  \caption{$C_{m,k}$ for 4 spans}
\end{subfigure}
\caption{perturbation coefficients contour plot at the truncation threshold of -25 dB  for fiber lengths of 1, 2, and 4 spans, respectively.}
\label{fig:perturbationcoeffMatrix}
\end{figure*}
\begin{table}[t]
\centering
    \caption{FIR filter and $\mathbf{c_0}$ lengths per step for PA-LDBP initialization.}
    \begin{tabular}{|P{2.5cm}|P{2.5cm}|P{1.5cm}|}\hline
      Scenarios (span(s)/step)   & FIR filter lengths  &  $\mathbf{c_0}$ lengths \\ \hline
      1      &  77 & 11 \\ \hline
       2   &   149 & 25 \\ \hline
       4     &  293 & 31\\ \hline
       10    &   725 &41 \\ \hline
    \end{tabular}
    \label{tab:FIRfilters}
\end{table}
\subsubsection{Configurations with Different Number of Spans per Step}
According to \eqref{perturbationcoeff} and \eqref{eq:spmxpm}, a given pulse interacts nonlinearly with a certain number of neighbouring pulses over a distance, and this number depends on the fiber length. To investigate the flexibility of the proposed scheme, we consider four scenarios with 1, 2, 4, 10 spans per step. Accordingly, 
$\mathbf{h}$ and $\mathbf{c_0}$ are initialized with fiber lengths of 80 km, 160 km, 320 km, and 800 km (note that same $\mathbf{h}$ and $\mathbf{c_0}$ are used to initialize each steps in the LDBP and PA-LDBP). For $\mathbf{h}$, the FIR filter is obtained based on the least square criterion \cite{hager2020physics}, and $\mathbf{c_0}$ can be computed by \eqref{perturbationcoeff}. The length of $\mathbf{c_0}$ depends on the window length corresponding to a truncation threshold \cite{orappanparasoman2021second}. The truncation is to keep only the significant coefficients for complexity reduction. A truncation criterion, defined as $20 \text{log}_{10} |C_{m,k}/C_{0,0}|=\chi$,  is applied, with $\chi$ as a truncation threshold. In Fig. \ref{fig:perturbationcoeffMatrix}, we show the values of $\chi$ ranging from $-25$~dB to $-5$~dB with a step size of 5 dB for different spans of fiber length. Given a selected truncation threshold, the numbers of significant terms of perturbation coefficients increases with augmenting the number of spans because of the dispersive effect. In the simulation, a truncation threshold of $-20$~dB is used, and the corresponding length of $\mathbf{c_0}$ for the four scenarios are 11, 25, 51, and 101, respectively. This threshold value is a good compromise between performance and complexity, as the performance improvement of PA-LDBP with a smaller threshold is insignificant. 
Furthermore, in the cases of 4 and 10 spans per step, we found that the lengths of $\mathbf{c_0}$ can be reduced to 31 and 41 after the training, respectively, without performance penalty. The lengths of the FIR filter and $\mathbf{c_0}$ to initialize PA-LDBP for the 4 scenarios are summarized in Table  \ref{tab:FIRfilters}.

\subsubsection{Training of PA-LDBP}
The PA-LDBP is implemented in TensorFlow. The input is the signal after coherent detection with 2 samples per symbol. The training data set includes 256 frames and each frame includes 2048 samples, while the testing data set includes 64 frames with the same frame length as the training data set. Therefore, the $Q^2$ factor is calculated from 393,216 bits. 
Adam is chosen as the optimizer with a learning rate of 0.001, and the batch size is 32. The loss function is the mean-squared error defined as $\mathcal{L}( {\mathbf{s}},\mathbf{\hat{s}})=\sum_{n=1}^{N} | {{s_n}}-{\hat{s}_n}|^2/N$, where $\hat{s}_n$ is the $n$th symbol after downsampling and phase de-rotation. Furthermore, the effective SNR, calculated by $10\log_{10}\left(\mathcal{L}( {\mathbf{s}},\mathbf{\hat{s}})^{-1}\right)$, is employed as an alternative performance metric to better demonstrate the DNN's convergence performance in the next section.

\begin{figure}[t]
\centering
\includegraphics[width=0.5\textwidth]{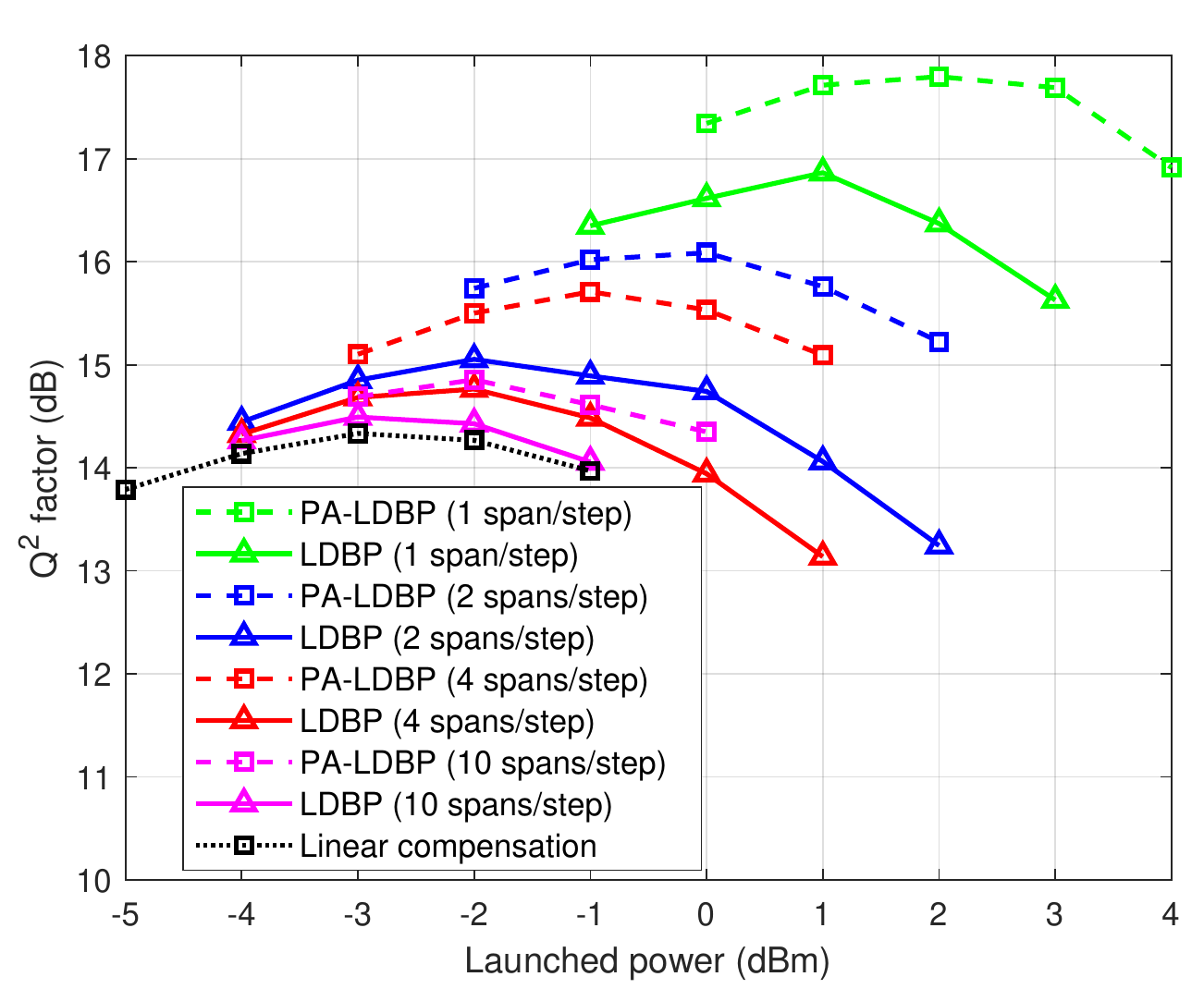}
\caption{Performance comparisons between LDBP and PA-LDBP.}
\label{fig:LPAvsLDBP}
\end{figure}

\subsection{Performance of the PA-LDBP}
First, we present the $Q^2$ factor performance of the proposed PA-LDBP and compare it to that of LDBP, as shown in Fig. \ref{fig:LPAvsLDBP}. The performance gains of PA-LDBP over linear compensation  are approximately 3.5 dB, 1.8 dB, 1.4 dB, and 0.6 dB, when the numbers of spans  per step are 1, 2, 4, 10, respectively. 
Furthermore, PA-LDBP outperforms LDBP by 0.9 dB, 1.1 dB, 0.9 dB, and 0.5 dB for the same number of spans per step, respectively. 
These performance improvements are expected, given the fact that PA-LDBP compensates for IXPM  along with SPM at each nonlinear step. Performance gains diminish  when the number of spans per step increases for both PA-LDBP and LDBP, as a large number of spans per step results in a numerical approximation error in calculating the  nonlinear distortion field. 
However, it is noticed that PA-LDBP with 10 spans per step achieves a similar gain when compared to LDBP with 2 spans per step.  The overall complexity with 10 spans per step  would be much lower than the case with 2 spans per step.  Details are shown in the complexity section. 

\vspace{-0.5cm}
\subsection{Initialization for PA-LDBP}
The initialization is critical to  DNN's  performance, as an inappropriate initialization could lead to slow learning or divergence.  The parameter initialization for the linear steps is similar to that of the LDBP, which has been discussed in \cite{hager2020physics}. Therefore, we focus on the discussion of the initialization for the nonlinear steps. In particular, for the four scenarios, the initial FIR filter length and number of perturbation coefficients are listed in Table \ref{tab:FIRfilters}. 
\begin{figure}[t]
\centering
\includegraphics[width=0.5\textwidth]{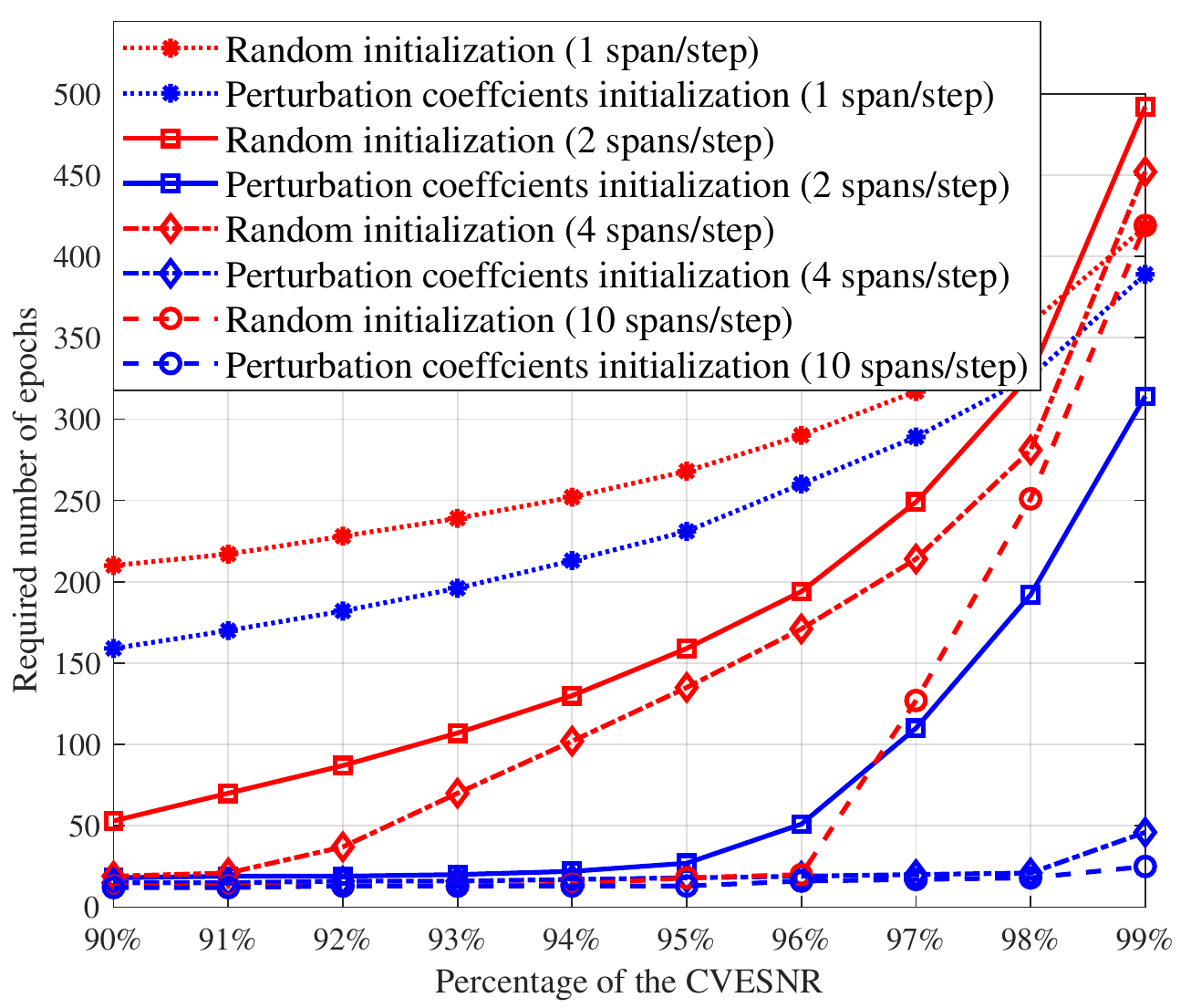}
\caption{Number of epochs required to achieve a certain level of the CVESNR with different initializations for PA-LDBP.}
\label{fig:InitImpact}
\end{figure}
Similar to what has been observed in \cite{hager2020physics} for the linear steps, it turns out that the initialization of nonlinear step in PA-LDBP plays an important role on the convergence of the DNN. 
Two initialization scenarios are considered: the initialization with the perturbation coefficients calculated by \eqref{perturbationcoeff} and random initialization. For the latter case, the real and imaginary parts of all filter taps are drawn from a Gaussian distribution with zero mean and unit variance. 
The results are shown in Fig. \ref{fig:InitImpact}, where the number of epochs to achieve 90\% to 99\% of the converged value of the effective SNR (CVESNR) is illustrated to indicate the quality and speed of convergence.  Note that when the random initialization is used, 20 realizations are performed to obtain the average. As can be seen, 

the initialization with the perturbation coefficients leads to a fast and good convergence of the DNN, especially when the number of spans per step is greater than one. This could be an advantage for an  elastic optical network where training is required more frequently to cope with the adaptive transmission.

\section{Complexity Analysis and Reduction}
Complexity is crucial when designing a digital NLC technique. The complexity metric adopted here is the number of real-valued multiplications, as they typically consume significantly more computation resources than other operations. In this section, the complexity of PA-LDBP is investigated first, followed by two proposed methods to reduce the complexity. Then, the performance versus complexity of PA-LDBP and LDBP is provided. 
\subsection{The Complexity of PA-LDBP}
Firstly, the complexity of LDBP is analysed as a benchmark. At each linear step, by considering that the FIR filter is symmetric,  the number of real-valued multiplications per sample is $4 \cdot \text{ceil}(N_{\text{CD}}/2)$, where $N_{\text{CD}}$ is the FIR filter length and $\text{ceil}(\cdot)$ is the ceiling function. At each nonlinear step, each sample is squared (2 real-valued multiplications), then multiplied by 
$\gamma$ (1 real-valued multiplication), followed by a phase rotation (4 real-valued multiplications). The exponential function is implemented by a look-up table.
The complexity analysis for PA-LDBP is done in a similar way, and the major difference lies  in the nonlinear steps. On top of the LDBP complexity in each nonlinear step, extra $4 \cdot \text{ceil}(N_{\text{PB}}/2)$ real-valued multiplications per sample are required, where $N_{\text{PB}}$ is the number of perturbation coefficients at each step.

\subsection{Complexity Reduction}

We propose two methods to reduce the overall complexity of PA-LDBP, as presented below.
 
\subsubsection{Perturbation Coefficient  Pruning}
\begin{figure*}[t!]
\centering
\begin{subfigure}{0.46\textwidth}
  \centering
  \includegraphics[width=1\linewidth]{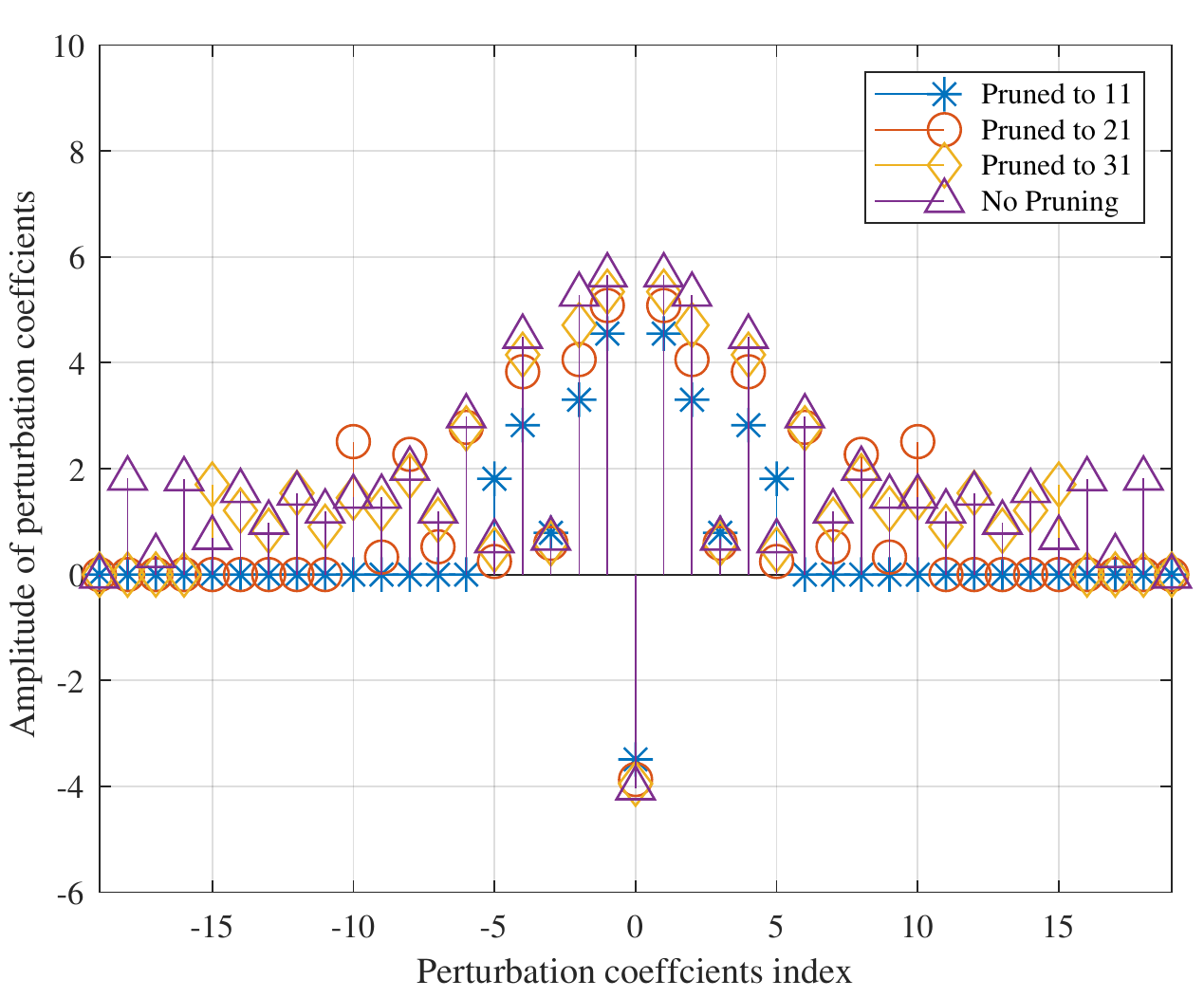}
  \caption{$C_{m,k}$ at Step 1}
\end{subfigure}%
\begin{subfigure}{0.46\textwidth}
  \centering
  \includegraphics[width=1\linewidth]{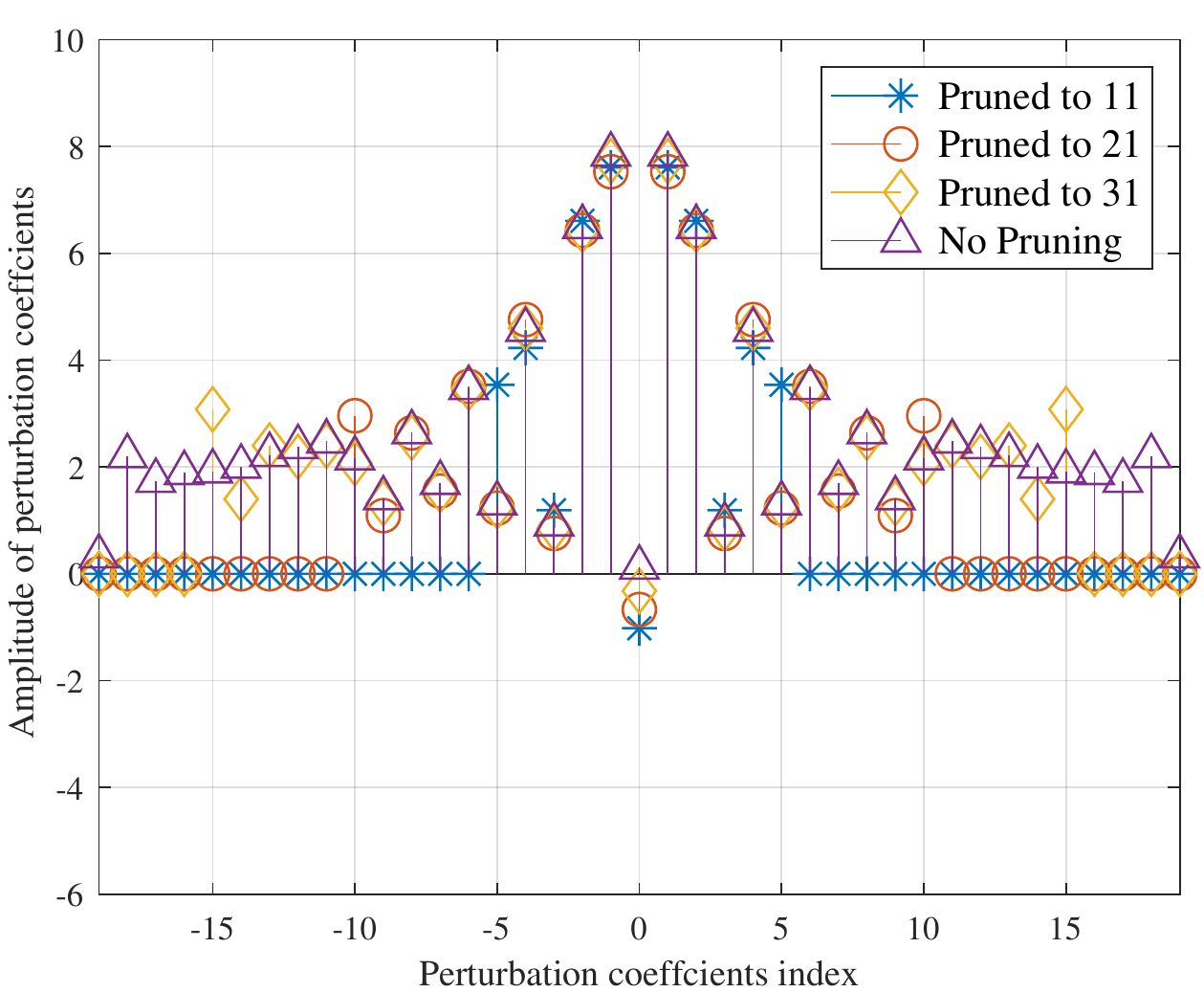}
  \caption{$C_{m,k}$ at Step 2}
\end{subfigure}%
\caption{Prior- and post-pruning perturbation coefficients at different steps, for 10 spans per step.}
\label{fig:perturbationcoeffPruning}
\vspace{-0.3cm}
\end{figure*}
Network  pruning is important for memory size and complexity reduction, and aims to remove redundant weights or neurons that do not contribute significantly to the accuracy of the networks output. In \cite{hager2020physics}, the CD FIR filter length can be decreased significantly through progressive pruning, which contributes a substantial complexity reduction with negligible performance penalty. More specifically, the lengths of the FIR filters after pruning are 37, 51,  95, 251 per step for the 4 scenarios, respectively. Compared to the values in Table \ref{tab:FIRfilters}, the FIR filter lengths are pruned more than half. 
This enhancement is attributed to the efforts of joint filter  design by the DNN: all the steps are incorporated into a multi-objective optimization problem instead of the standard least-square  formulation on one step.  
The pruning for linear steps from LDBP can be inherited by PA-LDBP. Furthermore, the trainable parameters in the nonlinear steps, i.e., the perturbation coefficients in $\mathbf{c_0}$,  can be pruned as well.

Illustrative examples are provided in Table \ref{tab:CompComp}, where the performance for 10 spans per step is shown with 31, 21, and 11 perturbation coefficients, respectively. Out of these cases, no performance penalty is observed when reducing the number of perturbation coefficients from 41 to 31. The amplitude of perturbation coefficients after pruning at each step  is shown for steps 1 and 2 in Fig. \ref{fig:perturbationcoeffPruning}. As can be seen, the coefficients at the two steps are not identical after training, while the symmetry of perturbation coefficients is preserved after pruning. 

Pruning the  perturbation coefficients is beneficial for complexity reduction  with a negligible performance penalty. The reason is similar to what has been stated for the pruning strategy for the FIR filter in LDBP, i.e., the efforts of joint filter  design by the DNN.

\begin{table}[t]
    \centering
        \caption{The performance with perturbation coefficients pruning (10 spans per step at launch power -2 dBm).}
    \begin{tabular}{|p{3.6cm}|c|c|c|c|}\hline
         Number of perturbation coefficient after pruning   & 31 &21 &11 \\ \hline
           $Q^2$-factor (dB)    & 14.99 &14.92 &14.82 \\ \hline 
    \end{tabular}
    \label{tab:CompComp}
    \vspace{-0.4cm}
\end{table}

\subsubsection{CD Compensation in Frequency Domain}

In \cite{hager2020physics,fan2020advancing}, convolution operations are embedded in the matrix operations in the DNN for CD compensation, and in \cite{sidelnikov2021advanced}, a CNN is employed instead. These TDEs that compensate for the CD at each linear step might be efficient when the FIR filter length is relatively short. More importantly, the filter coefficients pruning is straightforward to be implemented in time domain. However, for multiple spans per step, frequency domain equalization (FDE) can be more computationally-efficient because a longer filter is required.

At each linear step, one pair of fast Fourier transformation (FFT) and inverse FFT, and multiplications with the FDE filter coefficients at the corresponding discrete frequencies are required. For processing long sequences,  the overlap-and-add (OLA) method is used to segment the sequence into shorter-length blocks. Using FDE, the number of real-valued multiplications per sample is $4 \cdot [ 2 \cdot N_{\text{FFT}} \text{log}_2(N_{\text{FFT}})+ N_{\text{FFT}}]/(N_{\text{FFT}} - N_{\text{CD}})$, where $N_{\text{FFT}}$ is the FFT size in the OLA method. The complexity-optimal FFT sizes 
  corresponding to the four investigated scenarios (from 1 span per step to 10 spans per step) are 256, 512, 1024, and 2048, respectively. In practice, we recommend to obtain the pruned TDE filter weights through training and perform   FDE at the inference stage, as complexity at the inference stage is of more practical interest. 

\begin{figure}[t!]
\centering
\includegraphics[width=0.5\textwidth]{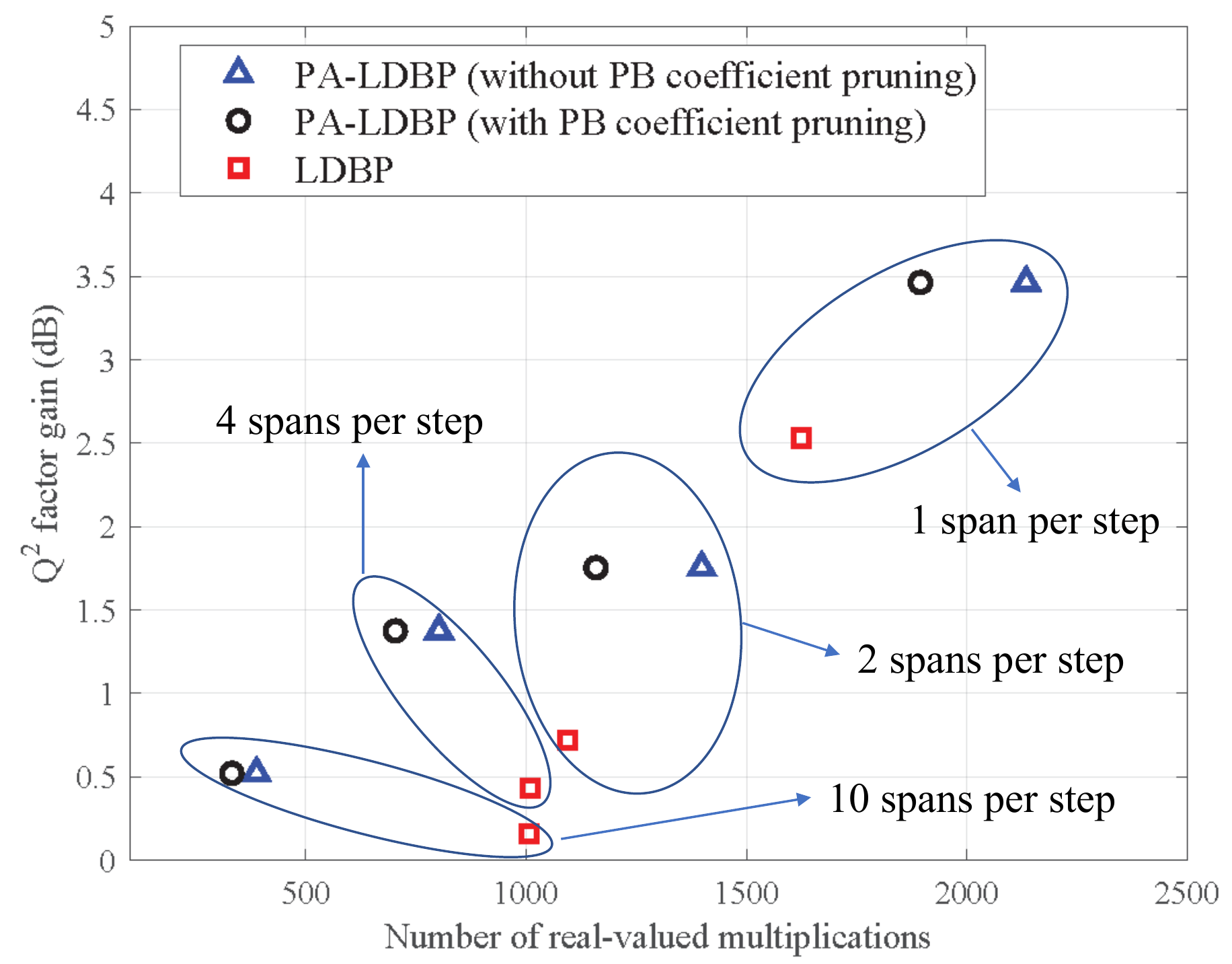}
\caption{Performance gain versus the number of real-valued multiplications for PA-LDBP and LDBP.}
\label{fig:GainvsComp}
\end{figure}

\subsection{Performance versus Complexity}

Figure \ref{fig:GainvsComp} shows the $\text{Q}^2$ factor gain  over linear compensation as a function of the number of real-valued multiplications per sample for LDBP  and FDE-based PA-LDBP. Pruning of the CD filter coefficients is applied in all cases. As can be seen, for 1 and 2 spans per step, PA-LDBP  without perturbation coefficients pruning  obtains improved performance at the price of higher complexity. 
However, when the number of spans per step increases, in the cases of 4 and 10 spans per step, PA-LDBP attains an enhanced performance with a reduced complexity.
Pruning for PA-LDBP reduces the number of perturbation coefficients to 5, 13, 21, and 27  per step for the 4 scenarios, respectively. 
We observe that this leads to a considerable reduction of complexity for 1 and 2 spans per step.  With the perturbation coefficients pruning, PA-LDBP bears a similar complexity as LDBP for 2 spans per step, while attaining an enhanced performance. 

\section{Conclusion}

A perturbation-aided machine learning scheme has been proposed for intra-channel nonlinearity compensation in coherent optical fiber communications systems. The proposed scheme improves the nonlinearity compensation at each nonlinear step by incorporating SPM and IXPM terms based on the first-order perturbation theory. This refinement has enabled a flexible restructuring of the existing LDBP technique. Furthermore, the CD compensation for multi-span per step can be performed in frequency domain, and this achieves a complexity reduction compared to the TDE-based LDBP. In addition, pruning of the perturbation coefficients has been investigated, and it has been showed that it successfully reduces the complexity. 
Overall, the proposed scheme attains  an enhanced performance when compared with the LDBP scheme,  with possibility of reduced complexity. Additionally, it provides practitioners with a flexible way to configure the compensation scheme with different numbers of spans per step.

\vspace{-0.3cm}

\bibliography{IEEEabrv,My_Ref_bib}
\bibliographystyle{IEEEtran}

\end{document}